\documentclass[onecolumn]{iau}

\usepackage{graphicx}% Include figure files
\usepackage{dcolumn}% Align table columns on decimal point
\usepackage{bm}% bold math
\usepackage[usenames]{color}
\definecolor{Blue}{rgb}{0,0.08,0.65}
\definecolor{Red}{rgb}{0.65,0.08,0.05}
\definecolor{Green}{rgb}{0.15,0.45,0.25}

\def \prd {{\it Phys. Rev. D}}

\def \apj {{\it ApJ}}

\def \mnras {{\it MNRAS}}

\title{ Non Gaussian Minkowski functionals and extrema counts for 2D sky maps % and galactic catalogues.
}

\author[D. Pogosyan, S. Codis, C. Pichon]
{Dmitri Pogosyan${}^{1,2,3}$ , Sandrine Codis${}^{2}$, 
Christophe Pichon${}^{2}$ }%
\affiliation{%
${}^{1}$ Department of Physics, University of Alberta, 11322-89 Avenue, Edmonton, Alberta, T6G 2G7, Canada\\
${}^{2}$ CNRS, UPMC, Institut d'astrophysique de Paris,  98 bis boulevard Arago, 75014, Paris, France \\
${}^{3}$ CNRS, Institut Lagrange de Paris, 98 bis boulevard Arago, 75014, Paris, France
}
\pubyear{2014}
\volume{308}  %% insert here IAU Symposium No.
\pagerange{xxx--xxx}
\jname{The Zeldovich Universe, Genesis and Growth of the Cosmic Web}
\editors{R. van de Weygaert, S. Shandarin, E. Saar \& J. Einasto, eds.}

\begin{document}
\maketitle

\begin{abstract}
In the conference presentation we have reviewed the theory of non-Gaussian
geometrical measures for the 3D Cosmic Web of the matter distribution 
in the Universe and 2D
sky data, such as Cosmic Microwave Background (CMB) maps that was developed
in a series of our papers.
The theory leverages symmetry of isotropic statistics
such as Minkowski functionals and extrema counts to develop post- Gaussian
expansion of the statistics in orthogonal polynomials of invariant descriptors
of the field, its first and second derivatives. The application of the
approach to 2D fields defined on a spherical sky was suggested, but never
rigorously developed.  In this paper we present such development treating
effects of the curvature and finiteness of the spherical space $S_2$ 
exactly,
without relying on the flat-sky approximation. We present Minkowski functionals,
including Euler characteristic and extrema counts 
to the first non-Gaussian correction, suitable for weakly non-Gaussian fields
on a sphere,
of which CMB is the prime example.
\end{abstract}

Random fields are ubiquitous phenomena in physics appearing in areas  ranging from
turbulence to the landscape of string theories.
In cosmology, the sky-maps of the polarized Cosmic Microwave Background (CMB) radiation 
-- a focal topic of current research --
is a prime example of  such  2D random fields, specified on $S_2$ spherical
space.
Modern view of the cosmos, developed primarily through statistical
analysis of these fields, points to a Universe that is 
statistically homogeneous and isotropic with a hierarchy of structures 
arising from small Gaussian fluctuations of quantum origin.
While the Gaussian limit provides the fundamental starting point in the study
of random fields \cite[]{Adler,Doroshkevich,BBKS},
non-Gaussian features of the CMB fields are of great interest.
Indeed, CMB inherits a high level of Gaussianity from 
initial fluctuations, but small non-Gaussian deviations may
provide a unique window into the details of processes in the early Universe.
The search for the best methods to analyze 
non-Gaussian random fields  is ongoing.  

In the paper \cite{PGP} the general invariant based formalism for computing
topological and geometrical 
characteristics of non Gaussian fields was presented. The general formula for
the Euler characteristic
to all orders has been derived, which encompasses the well known
first order correction of \cite{Matsubara}, 
and which was later confirmed to  the next order by \cite{matsu10}. 
This work was followed by the detailed exposition of the theory in 2D and 3D
flat (Cartesian) space in \cite{PPG} and \cite{Gay12},
and generalized to the 3D redshift
space where isotropy is broken in \cite{Codisetal}.

The goal of this paper is to extend these results to the fields
defined on a finite curved spherical space $S_2$ without reliance
on the flat field (small angle) approximation.
While these proceedings were being prepared,
similar work has been done for statistics of peaks in 
the Gaussian case within \cite{Marcos-Caballero}.
Here our focus is on non-Gaussian corrections.
We discuss how to compute
exact Minkowski functionals 
for the  excursion sets of a scalar field on a $S_2$ sphere
to all orders in non-Gaussian expansion and provide an explicit expression
for the Euler characteristic to first order.
Expressions for the total extrema counts to the
first non-Gaussian order are also given, while analytical formulas for
differential extrema counts to the same order will be published elsewhere
due to their length.
These results have a direct relevance to CMB data analysis.

\section{Joint distribution function of the field and its derivatives
on $S_2$ sphere.}

The statistics of Minkowski functionals, including the Euler number, as well
as extrema counts requires the knowledge of the one-point joint probability
distribution function (JPDF) $P(x,x_i,x_{ij})$
of the field $x$ (assumed to have zero mean), its
first, $x_i$, and second, $x_{ij}$, derivatives. 
Let us consider a random field $x$ defined on a 2D sphere $S_2$ of radius $R$
represented as the expansion in spherical harmonics
\begin{equation}
x(\theta,\phi) = \sum_{l=0}^\infty \sum_{m=-l}^l a_{lm} Y_{lm}(\theta,\phi)
\end{equation}
where for the Gaussian statistically homogeneous and isotropic field
random coefficients $a_{lm}$ are uncorrelated with $m$-independent
variances $C_l$ of each harmonic
\begin{equation}
\langle a_{lm} a^*_{l^\prime m^\prime}\rangle =
C_l \delta_{l l^\prime} \delta_{m m^\prime}
\end{equation} 
The variance of the field is then given by
\begin{equation}
\sigma^2 \equiv \left\langle x^2 \right\rangle =
\frac{1}{4\pi} \sum_l C_l (2l+1) 
\end{equation}

When considering derivatives in the curved space, we use covariant derivatives
$x_{;\theta}$, $x_{;\phi}$, ${x^{;\theta}}_{;\theta}$, ${x^{;\phi}}_{;\phi}$, 
${x^{;\theta}}_{;\phi}$ where it will be seen immediately that mixed version
for the second derivatives is the most appropriate choice.
The 2D rotation-invariant combinations of derivatives are
\begin{equation}
q^2 = x_{;\phi} x^{;\phi} + x_{;\theta} x^{;\theta} ~,~
J_1 = \left(x^{;\theta}_{;\theta}+x^{;\phi}_{;\phi}\right)^2 ~,~
J_2 = \left(x^{;\theta}_{;\theta}-x^{;\phi}_{;\phi}\right)^2
+ 4 x^{;\theta}_{;\phi} x^{;\phi}_{;\theta} 
\end{equation}
where $J_1$ is linear in the field and $q^2$ and $J_2$ are quadratic,
always positive, quantities.
The derivatives 
are also random Gaussian variables, which variances are easily computed
\begin{eqnarray}
\sigma_1^2 &\equiv& \langle q^2 \rangle =
\frac{1}{4\pi R^2} \sum_l C_l l (l+1) (2l+1) 
\\
\sigma_2^2 &\equiv& \langle J_1^2 \rangle =
\frac{1}{4\pi R^4} \sum_l C_l l^2 (l+1)^2 (2l+1) \\
\sigma_2^{\prime 2} &\equiv&  \langle J_2 \rangle =
\frac{1}{4 \pi R^4} \sum_l C_l (l-1) l  (l+1) (l+2) (2l+1)
\end{eqnarray}
where the fundamental difference between a sphere and the 2D Cartesian space
is in the fact that $\sigma_2^\prime \ne \sigma_2$.
Among the cross-correlations the only non-zero one is between
the field and its Laplacian
$\left\langle x \left(x^{;\theta}_{;\theta}+x^{;\phi}_{;\phi}\right)\right\rangle = - \sigma_1^2 $.

From now on we rescale all random quantities by their variances, so that
rescaled variables have $\langle x^2\rangle = \langle J_1^2 \rangle 
= \langle q^2 \rangle = \langle J_2 \rangle =1$.
Introducing  $\zeta=(x+\gamma
J_1)/\sqrt{1-\gamma^2}$ (where the spectral parameter $\gamma=
- \langle x J_1 \rangle = \sigma_1^2/(\sigma\sigma_2)$)
leads to the following simple JPDF for the Gaussian 2D fields
\begin{equation}
G_{\rm 2D} = \frac{1}{2 \pi} 
\exp\left[-\frac{1}{2} \zeta^2 - q^2 - \frac{1}{2} J_1^2 - J_2 \right] \,.
\label{eq:2DG}
\end{equation}

In \cite{PGP} we have observed that for non-Gaussian  JPDF the invariant
approach immediately suggests a Gram-Charlier expansion  in
terms of the orthogonal polynomials defined by the kernel $G_{\rm 2D}$.
Since $\zeta$, $q^2$, $J_1$ and $J_2$ are uncorrelated variables in the
Gaussian limit,
the resulting expansion is
\begin{eqnarray}
P_{\rm 2D}(\zeta, q^2, J_1, J_2) &=&  G_{\rm 2D} \left[
\vphantom{ \frac{(-1)^{j+l}}{i!\;j!\; k!\; l!}}
1 +  \right.
\nonumber \\
\sum_{n=3}^\infty \sum_{i,j,k,l=0}^{i+2 j+k+2 l=n}  && \left.
\frac{(-1)^{j+l}}{i!\;j!\; k!\; l!} 
\left\langle \zeta^i {q^2}^j {J_1}^k {J_2}^l \right\rangle_{\rm GC}
H_i\left(\zeta\right) L_j\left(q^2\right)
H_k\left(J_1\right) L_l\left(J_2\right)
\right],
\label{eq:2DP_general}
\end{eqnarray}
where terms are sorted in the order of the field power $n$
and $\sum_{i,j,k,l=0}^{i+2 j+k+2 l=n} $
stands for summation over all combinations
of non-negative $i,j,k,l$ such that $i+2j+k+2l$ adds 
to the order of the expansion term $n$. $H_i$ are ({\it probabilists'}) Hermite
and $L_j$ are Laguerre polynomials.
The coefficients of expansion 
\begin{equation}
\left\langle \zeta^i {q^2}^j J_1^k J_2^l \right\rangle_{\scriptscriptstyle{\mathrm{GC}}} \!\! =
\frac{j! \; l!}{(-1)^{j+l}} 
\left\langle \vphantom{ \zeta^i {q^2}^j J_1^k J_2^l } \!
H_i\left(\zeta\right) L_j\left(q^2\right)
H_k\left(J_1\right) L_l\left(J_2\right) \!\right\rangle.
\label{eq:GCtomoments2D}
\end{equation}
are related (and for the first non-Gaussian order $n=3$ are equal) 
to the moments of the field and its derivatives (see \cite{Gay12} for details).

Up to now our considerations are 
practically identical to the theory in the Cartesian space,
which facilitates using many of the Cartesian calculations.
We stress again the only, but important, difference being 
$\sigma_2^\prime \ne \sigma_2$.  We shall see in the next sections
how this difference plays out. Here we introduce the spectral parameter
$\beta$ that describes this difference
\begin{equation}
\beta \equiv 1 - \frac{ \sigma_2^{\prime 2}}{ \sigma_2^2}
=  2 \frac{ \sum_l C_l l (l+1) (2l+1) }{ \sum_l C_l l^2 (l+1)^2 (2l+1)}
\label{eq:beta}
\end{equation}

Let us review the scales and parameters that the theory has.
As in the flat space, we have two scales $R_0 = \sigma/\sigma_1$ and
$R_* = \sigma_1/\sigma_2$ and the spectral parameter $\gamma = R_*/R_0$
(which also describes correlation between the field and its 
second derivatives). On a sphere we have a third scale, the curvature
radius $R$.
The meaning of the additional spectral parameter 
$\beta$ becomes clear if we notice that
$\sigma_2^2 - \sigma_2^{\prime 2} = 2 \sigma_1^2/R^2$, thus
$\beta = 2 R_*^2/R^2$, i.e describes the ratio of the 
correlation scale $R_*$
to the curvature of the sphere. As with $\gamma$, $\beta$ varies from $0$
to $1$, with $\beta=0$ corresponding to the flat space limit.
From Eq.~(\ref{eq:beta}) we find that $\beta=1$ is achieved when the
field has only the monopole and the dipole in its spectral decomposition.

\section{Minkowski functionals on $S_2$ beyond the Gaussian limit}

There are three Minkowski functionals that are defined for the
excursion set above threshold $\nu$ of a 2D field, 
namely the filling factor, $f_V(\nu)$, i.e
the volume fraction occupied by the region above the threshold $\nu$,
the length (per unit volume) of isofield contours, ${\cal L}(\nu)$ and
Euler characteristic $\chi(\nu)$. Statistics of the first two
do not depend on the second derivatives of the field, and thus
are identical on $S_2$ and the 2D Cartesian space. Here, for completeness,
we reproduce
the non-Gaussian expansions for these quantities from \cite{Gay12} 
\begin{equation}
 f(\nu) = 
 \frac{1}{2} \mathrm{Erfc} \left(\frac{\nu}{\sqrt{2}}\right) +
\frac{1}{\sqrt{2 \pi}}e^{-\frac{\nu^2}{2}}
\sum_{n=3}^\infty \frac{ \langle x^n \rangle_\mathrm{{\scriptscriptstyle GC}} }{n!} H_{n-1}(\nu).\label{eq:defF}
\end{equation}

\begin{equation}
 {\cal L}(\nu)=\frac{1}{2\sqrt{2} R_0} e^{-\frac{\nu^2}{2}} 
\left( 1 + \frac{1}{2\sqrt{\pi}} \sum_{n=3}^{\infty} \sum_{i,j}^{i+2j=n} \frac{(-1)^{j+1}}{i! j!}  \frac{\Gamma(j-\frac{1}{2})}{\Gamma(j+1)} \left\langle x^i q^{2j} \right\rangle_\mathrm{{\scriptscriptstyle GC}} H_i(\nu) \right)\,.\label{eq:defCont2Dfinal}
\end{equation}

Euler characteristic density of the region above a threshold $x=\nu$
is a more interesting case. It is given by the average
of the determinant of the Hessian matrix of the second derivatives
of the field at the points
where the first derivatives vanish \cite{Adler,Longuet}
\begin{equation}
\chi(\nu) = \int_\nu^\infty \!\! \mathrm{d} x 
\int {\rm d}^3 x_{ij} P(x,x_i=0,x_{ij})
\det(x_{ij}) \,.
\label{eq:chi_int}
\end{equation}
It has been argued in \cite{PGP} that on $S_2$ the determinant should
be that of the Hessian of the mixed covariant derivatives
$\det({x^{;i}}_{;j})$.  It is this choice that provides the density relative
to the
invariant volume element $R^2 \sin^2\theta \mathrm{d}\theta \mathrm{d}\phi$
%\footnote{if one uses $\det(x_{;i;j})$ one obtains density in
%the coordinate volume $\mathrm{d}\theta\mathrm{d}\phi$}
and has a scalar trace equal to the Laplacian of the field.
Using scaled invariant variables
\begin{equation}
\det({x^{;i}}_{;j}) = \frac{\sigma_2^2}{4} \left(J_1^2-(1-\beta) J_2 \right)
\equiv \sigma_2^2 I_2 ~,
\end{equation}
where we have introduced another scaled quadratic invariant $I_2$.
In terms of the eigenvalues of the Hessian, 
$\sigma_2^2 I_2 = \lambda_1 \lambda_2$,
while $\sigma_2^2 J_1 = \lambda_1 + \lambda_2$ and
$\sigma_2^2 (1-\beta) J_2 = (\lambda_1 - \lambda_2)^2$.

In the Gaussian limit the Euler characteristic density becomes
%\begin{equation}
%\chi(\nu) = \frac{\sigma_2^2}{8 \pi^2 \sigma_1^2 \sqrt{1-\gamma^2}}
%\int_{\nu}^\infty \!\! \mathrm{d} x 
%\int_{-\infty}^\infty \!  {\rm d} J_1 \int_0^\infty \! {\rm d} J_2
%\exp\left[-\frac{1}{2} \frac{(x + \gamma J_1)^2}{1-\gamma^2}
%- \frac{1}{2} J_1^2 - J_2\right]
%\left( J_1^2- (1-\beta) J_2 \right)
%\end{equation}
\begin{equation}
\chi(\nu) = \frac{\sigma_2^2}{8 \pi^2 \sigma_1^2}
\int_{-\infty}^\infty \!  {\rm d} J_1 \int_0^\infty \! {\rm d} J_2
\int_{\frac{\nu+\gamma J_1}{\sqrt{1-\gamma^2}}}^\infty \!\! \mathrm{d} \zeta 
\exp\left[-\frac{1}{2} \zeta^2 - \frac{1}{2} J_1^2 - J_2\right]
\left( J_1^2- (1-\beta) J_2 \right) 
\label{eq:chi_gauss}
\end{equation}
It evaluates to 
\begin{equation}
\chi(\nu) = \frac{\gamma^2}{4 \pi \sqrt{2 \pi} R_*^2} \nu e^{-\frac{\nu^2}{2}}
 + \frac{\beta}{8 \pi R_*^2} \mathrm{erfc} \left(\frac{\nu}{\sqrt{2}}\right)
\end{equation}
which differ from the well known Cartesian result by the $\beta \ne 0$ term.
On a sphere which has a finite volume $4 \pi R^2$ it is appropriate to quote
the total Euler characteristic in the whole volume, which, recalling the
relation between $\gamma$, $\beta$, $R$ and $R_*$ becomes
\begin{equation}
4 \pi R^2 \chi(\nu) = 
\frac{R^2}{\sqrt{2 \pi} R_0^2} \nu e^{-\frac{\nu^2}{2}}
 + \mathrm{Erfc} \left(\frac{\nu}{\sqrt{2}}\right)
\end{equation}
which explicitly demonstrates that if $\nu=-\infty$. i.e the whole
space is included in the excursion set, the total Euler characteristic is
equal to that of a sphere, $4 \pi R^2 \chi(-\infty) = 2$, as expected.

Evaluation of the non-Gaussian expansion for $\chi(\nu)$ 
entails integrating Eq.~(\ref{eq:chi_int}) with the distribution function
$P_{2D}$ given by Eq.~(\ref{eq:2DP_general}). The procedure is similar
to that in Cartesian space as elaborated in detail in \cite{Gay12}
and which led to the complete expression 
for the Euler characteristic to all orders first reported in
\cite{PGP}.  Indeed, the quantity $I_2$ that we then average over the whole
range of $J_2$ can be rewritten as
$H_2(J_1) + \beta  + (1-\beta) L_1(J_2)$. Thus only the $l=0,1$ terms
of the expansion, i.e containing $L_0(J_2)$ or $L_1(J_2)$, do not vanish
after integration. Here we should limit ourselves to presenting only
the result of the most practical use - up to the first, cubic in the field,
non-Gaussian correction 
\begin{eqnarray}
\lefteqn{\chi (\nu) = \frac{\beta}{8 \pi R_*^2}
\mathrm{Erfc} \left(\frac{\nu}{\sqrt{2}} \right)
+ \frac{1}{4 \pi \sqrt{2 \pi} R_*^2}  
\exp\left(-\frac{\nu^2}{2}\right) } 
\nonumber \\
&& \times \left[ 
\gamma^2 H_1 (\nu) + 2 \gamma  \left\langle q^2 J_1\right\rangle +4 \left\langle x  I_2\right\rangle -\left( \gamma ^2 \left\langle x q^2 \right\rangle + \gamma  \left\langle x ^2 J_1\right\rangle \right) H_2(\nu)+\frac{\gamma^2}{6} \left\langle x ^3\right\rangle  H_4(\nu) 
\right. \nonumber \\
&& \quad\quad + \left. \beta \left(
- \langle x q^2 \rangle H_0(\nu) + 
\frac{1}{6} \langle x^3 \rangle H_2(\nu)
\right)
\right]\,,
\label{eq:euler}
\end{eqnarray}
where the Gram-Charlier moments of the non-primary
variables $x$ and $I_2$ are understood as correspondent combinations
of Gram-Charlier moments of the expansion variables $\zeta$, $J_1$ and $J_2$.
The first term of Eq.~(\ref{eq:euler}) is the Gaussian result on
the sphere that is responsible for the total Euler number of the excursion
set to be that of the total sphere when $\nu=-\infty$. The last 
$\propto \beta$ terms is a correction to the non-Gaussian result due to
the curvature of the sphere. In conclusion we as well write explicitly
the result for the total Euler number above threshold $\nu$
\begin{eqnarray}
\lefteqn{4 \pi R^2 \chi (\nu) = 
\mathrm{Erfc} \left(\frac{\nu}{\sqrt{2}} \right)
+ \frac{2}{\sqrt{2 \pi}}  
\exp\left(-\frac{\nu^2}{2}\right) 
\left(
\frac{1}{6} \langle x^3 \rangle H_2(\nu)
- \langle x q^2 \rangle H_0(\nu) 
\right)
}
\nonumber \\
&& + \frac{R^2}{\sqrt{2 \pi} R_0^2}  
\exp\left(-\frac{\nu^2}{2}\right) 
\\
&& \times \left[ 
H_1 (\nu) + \frac{2}{\gamma}  \left\langle q^2 J_1\right\rangle + \frac{4}{\gamma^2} \left\langle x  I_2\right\rangle -\left(\left\langle x q^2 \right\rangle + \frac{1}{\gamma}  \left\langle x ^2 J_1\right\rangle \right) H_2(\nu)+\frac{1}{6} \left\langle x ^3\right\rangle  H_4(\nu) 
\right]\,.
\nonumber
\label{eq:euler_tot}
\end{eqnarray}

\section{Extrema counts on $S_2$ beyond the Gaussian limit}
The number density of extrema above a threshold $\nu$
is given by an integral very similar to the Euler characteristic
(\cite{Adler,Longuet})
\begin{equation}
n_{\rm ext}(\nu) = \int_\nu^\infty dx \int {\rm d}^3 x_{ij}
P(x,x_i=0,x_{ij})
|x_{ij}| \Theta_{ext}(\lambda_m)\,.
\label{eq:ext_int}
\end{equation}
where the function $\Theta_{ext}(\lambda_m)$ chooses the regions
of integration in the space
of second derivatives with appropriate to the particular extremum type
signs of the Hessian eigenvalues. In 2D, assuming $\lambda_1 \ge \lambda_2$,
$\Theta_{ext}(\lambda_m) = \Theta(-\lambda_1)$ for maxima, $=\Theta(\lambda_2)$ for
minima, and $=\Theta(\lambda_1)\Theta(-\lambda_2)$ for saddle points. 
%$\Theta$ being the Heaviside function.
In particular, a well-known topological relation gives
\begin{equation}
\chi(\nu)  = n_\mathrm{max}(\nu) - n_\mathrm{sad}(\nu) + n_\mathrm{min}(\nu)
\end{equation}

The integral Eq.~(\ref{eq:ext_int}) has a very transparent form when
the Hessian is described in invariant variables.  It is equivalent to 
Eq.~(\ref{eq:chi_gauss}), except that the limits of integration over $J_1$ 
are partitioned into the regions of fixed sign of the determinant $I_2$.
Namely, maxima correspond to the range 
$J_1 \in (-\infty, -\sqrt{(1-\beta)J_2})$, minima to 
$J_1 \in (\sqrt{(1-\beta)J_2},\infty)$ and saddle points to
$J_1 \in (-\sqrt{(1-\beta)J_2}, \sqrt{(1-\beta)J_2})$.

Calculations for the differential density of extremal points,
$\partial n_\mathrm{ext}/\partial(\nu)$,
can be carried out analytically even for the
general expression Eq.~(\ref{eq:2DP_general}) (see discussion for the flat case 
in \cite{Gay12}). The resulting expressions are cumbersome, and
here we limit ourselves to presenting results
for the total density of extrema to first non-Gaussian order only.
The total number density of maxima is given by
\begin{equation}
n_\mathrm{max} = \frac{\sigma_2^2}{4\sigma_1^2}
\int_0^\infty \! {\rm d} J_2
\int_{-\infty}^{-\sqrt{(1-\beta)J_2}} \!  {\rm d} J_1
P_\mathrm{2D}(q^2=0,J_1,J_2)
\left| J_1^2- (1-\beta) J_2 \right| 
\label{eq:ext_total}
\end{equation}
and, similarly, for the minima and the saddle points. The result
is
\begin{eqnarray}
n_{\rm max/min} &=&  \frac{(1-\beta)^{3/2}+\beta \sqrt{3-\beta}}
{8 \pi \sqrt{3-\beta} {R_*}^2} \pm
\frac{6(3-\beta) \left\langle q^2 J_1 \right\rangle 
- (5 - 3\beta) \left\langle J_1^3 \right\rangle
+ 6 (1-\beta) \left\langle J_1 J_2 \right\rangle}
{6 \pi \sqrt{2\pi} {R_*}^2 (3 - \beta)^2}
\,, \nonumber \\
n_{\rm sad} &=& \frac{(1-\beta)^{3/2}}{4 \pi \sqrt{3-\beta} {R_*}^2} 
\,,
\end{eqnarray}
where we immediately see that
$n_\mathrm{max}+n_\mathrm{min}-n_\mathrm{sad} = 
\beta/(4 \pi R_*^2)=2/(4 \pi R^2)$ as expected.
The total number 
of saddles, as well as of all the extremal points, $n_{\rm max}
+ n_{\rm min} + n_{\rm sad}$, are
preserved at first order (the latter following from the former),
but the symmetry between the minima and the maxima is  broken.

It is instructive to look how the Gaussian extrema counts are modified
by the properties of spherical space when the curvature radius is large
relative to the typical extrema separation scale $R_*$, i.e when $\beta$
is small. Up to first order in $\beta$
\begin{eqnarray}
n_\mathrm{max/min} &\sim& \frac{1}{8 \sqrt{3} \pi R_*^2} 
+ \frac{9-4\sqrt{3}}{36 \pi R^2}
\approx \frac{1}{8 \sqrt{3} \pi R_*^2} \left(1 + 0.8 R_*^2/R^2 \right)
\\
n_\mathrm{sad} &\sim& 
\frac{1}{4 \sqrt{3} \pi R_*^2} 
- \frac{2}{3 \sqrt{3} \pi R^2}
\approx \frac{1}{4 \sqrt{3} \pi R_*^2} \left(1 - 2.7 R_*^2/R^2 \right)
\end{eqnarray}
This shows that being on a sphere increases the number density of maxima
and minima, but decreases (and in a more significant way) the number of saddles.
Incidently, assuming large-angle CMB power spectrum, truncated at $l=30$ gives
$\beta \approx 1/170$, i.e 1\% correction to the count of extrema relative to
the flat-sky approximation.

 %%%%%%%%%%%%%
%\bibliography{proceedings}
%\bibliographystyle{plain}

\end{document}